\documentclass{jetpl}
\usepackage{epsfig,cite,color}
\twocolumn
\title{Exclusive open-charm near-threshold cross sections in a coupled-channel approach}
\rtitle{Exclusive open-charm near-threshold cross sections in a coupled-channel approach}

\author{T.\,V.\,Uglov}
\address{Lebedev Physics Institute, 119991, Leninsky prospect 53, Moscow, Russia\\
National Research Nuclear University MEPhI, 115409, Kashirskoe highway 31, Moscow, Russia\\
Moscow Institute of Physics and Technology, 141700, Institutsky lane 9, Dolgoprudny, Moscow Region, Russia}

\author{Yu.\,S.\,Kalashnikova}
\address{Institute of Theoretical and Experimental Physics, 117218, B.Cheremushkinskaya 25, Moscow, Russia\\
National Research Nuclear University MEPhI, 115409, Kashirskoe highway 31, Moscow, Russia}

\author{A.\,V.\,Nefediev}

\address{Institute for Theoretical and Experimental Physics, 117218, B.Cheremushkinskaya 25, Moscow, Russia\\
National Research Nuclear University MEPhI, 115409, Kashirskoe highway 31, Moscow, Russia\\
Moscow Institute of Physics and Technology, 141700, Institutsky lane 9, Dolgoprudny, Moscow Region, Russia}

\author{G.\,V.\,Pakhlova, P.\,N.\,Pakhlov}

\address{Lebedev Physics Institute, 119991, Leninsky prospect 53, Moscow, Russia\\
National Research Nuclear University MEPhI, 115409, Kashirskoe highway 31, Moscow, Russia\\
Moscow Institute of Physics and Technology, 141700, Institutsky lane 9, Dolgoprudny, Moscow Region, Russia}

\rauthor{ T.\,V.\,Uglov, Yu.\,S.\,Kalashnikova, A.\,V.\,Nefediev, G.\,V.\,Pakhlova, P.\,N.\,Pakhlov}

\abstract{Data on open-charm channels collected by the Belle Collaboration are analysed simultaneously using a unitary approach
based on a coupled-channel model in a wide energy range $\sqrt{s}=3.7\div 4.7$~GeV. The resulting fit provides a remarkably
good overall description of the line shapes in all studied channels. Parameters of 5 vector charmonium resonances are extracted
from the fit.}


\begin{document}
\maketitle

\setcounter{figure}{0}
\setcounter{table}{0}

\section{Introduction}

Four vector charmonia above the open-charm threshold, $\psi(3770)$,
$\psi(4040)$, $\psi(4160)$, $\psi(4415)$, were discovered forty years
ago in the $e^+e^-$ annihilation as peaks in the total hadronic cross
section~\cite{MarkI:cs_4415,MarkI:cs_3770,DASP:cs,DELCO:cs_3770,MarkII:cs_3770}.
Only thirty years later parameters of these states were updated from a
naive combined fit~\cite{seth:fit} for the Crystal Ball~\cite{cb:cs}
and the BES data~\cite{bes:cs}. In 2008 BES fitted the total hadronic
cross section (in terms of the ratio $R$)~\cite{bes:cs} taking into
account the interference and the relative phases between the exclusive
decays of the $\psi$-resonances~\cite{bes:fit}. As BES used model
predictions for the $\psi$'s decays into the two-body charmed final
states, the obtained parameters remain model-dependent. In its study
BES did not try to extract parameters of the vector charmoniumlike
states $Y(4008)$, $Y(4260)$, $Y(4360)$, and $Y(4660)$ observed since
2005 in the $e^+e^-$ annihilation by
BaBar~\cite{babar:y4260,babar:y4260_08,babar:y4360} and
Belle~\cite{belle:y4260:abe,belle:y4260:yuan,belle:y4360,Wang:2014hta,Liu:2013dau}
in the dipion transitions to light charmonia with the initial state
radiation (ISR). Except for the $Y(4008)$ resonance found by Belle
only in the $J/\psi \pi^+\pi^-$ final state, the $Y(4260)$, decaying
into $J/\psi \pi^+\pi^-$, as well as the $Y(4360)$ and $Y(4660)$,
decaying into $\psi(2S)\pi^+\pi^-$, are reliably established today. It
has to be noticed that the $Y$ states lying above the open-charm
threshold do not appear explicitly as peaks either in the total
hadronic cross section or in the exclusive $e^+e^-$ cross sections to the
open-charm final states measured later (the only vector charmoniumlike
state which reveals itself as a peak at threshold is the $X(4630)$
observed in the $\Lambda_c^+\Lambda_c^-$ final
state~\cite{belle:lala}.) It cannot be excluded however that some of
the $Y$ states could manifest themselves as coupled-channel effects
predicted in \cite{cleo:eichten}.

A comprehensive study of the exclusive $e^+e^-$ cross sections to
various open-charm final states could help one to extract parameters
of the $\psi$ states in a model-independent way and therefore to shed
light on the nature of the $Y$ family. Such cross sections were first
measured by
Belle~\cite{belle:dd,belle:ddst,belle:ddpi,belle:lala,belle:ddstpi,belle:dsds}
and BaBar~\cite{babar:dd,babar:ddst,babar:dsds} at the $B$-factories,
using the ISR in a wide energy region $\sqrt{s}=3.7\div 5.0$~GeV, and by
CLEO~\cite{cleo:cs}, using the energy scan over
$\sqrt{s}=3.97\div 4.26$~GeV at the charm factory. BaBar performed fits to
the measured two-body $D\bar{D}$, $D\bar{D}^*$, $D^*\bar{D}^*$, and $D_s^{(*)+}D_s^{(*)-}$
cross sections with the parameters of the $\psi$ states fixed to the
Particle Data Group (PDG)~\cite{Agashe:2014kda} values. From this
study only the ratios of the branching fractions for the $\psi(4040)$,
$\psi(4160)$, and $\psi(4415)$ decays to the $D\bar{D}$, $D\bar{D}^*$, and
$D^*\bar{D}^*$ were extracted. CLEO compared the $e^+e^-$ cross sections to
the same open-charm meson pairs with the predictions of an updated
potential model~\cite{cleo:eichten} that appeared to fail to describe
the data. Belle presented not only the two-body but also the
three-body $D^0 D^{(*)-} \pi^+$ and the charmed baryon
$\Lambda_c^+\Lambda_c^-$ cross sections and demonstrated that the sum
of the measured partial cross sections almost saturates the total
hadronic cross section.

Many attempts were made to describe the measured exclusive cross
sections
\cite{Li:2009pw,Cao:2014vca,Cao:2014qna,Limphirat:2013jga,Achasov:2012ss,Achasov:2013lwk,Zhang:2009gy,Zhang:2010zv,Segovia:2011zza,Segovia:2013wma,
vanBeveren:2010mg}.
Most of the authors
\cite{Li:2009pw,Cao:2014vca,Cao:2014qna,Limphirat:2013jga,Achasov:2012ss,Achasov:2013lwk,Zhang:2009gy,Zhang:2010zv}
were interested in the line shape of the $\psi(3770)$ in the $e^+e^-\to D\bar D$ reaction while others tried to describe the
$\psi(4415)$~\cite{Segovia:2011zza,Segovia:2013wma} or to extract the
parameters of higher charmonia
\cite{vanBeveren:2010mg}.
Although coupled-channel effects were taken into account by some
authors~\cite{Zhang:2010zv,Segovia:2011zza,Segovia:2013wma}, no
simultaneous fits of all measured exclusive cross sections have been
performed so far.

In this paper we present a coupled-channel fit to the exclusive cross
sections of the $e^+e^-$ annihilation to various open-charm final
states. The most accurate and coherent measurements in the full
$\sqrt{s}$ interval below 5~GeV for numerous two-body final states in
the $e^+e^-$ annihilation was presented by the Belle
Collaboration. The BaBar results are in a good agreement with the Belle
measurements but they are incomplete for our purpose and are of
a slightly worse accuracy. BES and CLEO provide measurements in several
points of their energy scan which does not allow one to trace all
features of the cross section in the full interval. We therefore stick
to the Belle data only, available at the Durham database~\cite{durh}.

As was demonstrated by Belle, the sum of the four channels ($D\bar{D}$,
$D\bar{D}^*$, $D^*\bar{D}^*$, $D\bar{D}\pi$) well saturates the inclusive hadronic cross
section in the region $\sqrt{s}\lesssim 4.7$~GeV, so we confine
ourselves by considering these four channels and this energy region only. Since the charged
and the neutral modes in each channel are related to each other by the
isospin symmetry which is a very accurate symmetry of QCD, we only
distinguish between the $D^+D^-$ and $D^0\bar{D}^0$ modes explicitly
since they are presented separately by Belle. For the other final
states only the data for the charged mode are available whose
contributions are therefore doubled to mimic the presence of the
neutral modes.

Then, for the $D\bar{D}\pi$ final state, Belle presented the cross section
for the $D^0 D^-\pi^+$ mode and demonstrated that this final state is
dominated by the contribution from the two-body mode $D\bar{D}_2$. To
convert the measured $D^0 D^-\pi^+$ cross section into the $D\bar{D}_2$ one
we correct the former by the ratio of the branching fractions
${\cal B}(D_2\to D\pi)/({\cal B}(D_2\to D\pi)+{\cal B}(D_2\to D^*\pi))$ \cite{Agashe:2014kda}.

Finally, when identifying the open-charm channels it should be taken
into account that 3 different final states are allowed for the
$D^*\bar{D}^*$ channels, namely, the $P$ wave with $S=0$, the $P$ wave
with $S=2$, and the $F$ wave with $S=2$. For convenience and in order
to avoid confusion we count such modes as independent channels. We
therefore stick to the set of 16 channels, thus counting the charge- and isospin-conjugated final states as independent ones,
\begin{equation}
\begin{array}{lcl}
D\bar{D},&\hspace*{3mm}&\mbox{2 channels},\\
D\bar{D}^*,&&\mbox{4 channels},\\
D_2\bar{D},&&\mbox{4 channels},\\[0mm]
[D^*\bar{D}^*]^P_{S=0},&&\mbox{2 channels},\\[0mm]
[D^*\bar{D}^*]^P_{S=2},&&\mbox{2 channels},\\[0mm]
[D^*\bar{D}^*]^F_{S=2},&&\mbox{2 channels}.
\end{array}
\label{channels}
\end{equation}
In what follows these channels are 
labelled by latin letters $i$, $j$, and so on.
Once the exact isospin limit is assumed, all parameters (but the charged and
neutral meson masses!) in the isospin-cojugated channels are taken equal to each
other.

Reactions $e^+ e^- \to D^{(*)}\bar{D}^{(*)}$ studied in this work are
expected to proceed through the 5 intermediate vector resonances,
\begin{equation}
\psi(2S),~\psi(3770),~\psi(4040),~\psi(4160),~\psi(4415),
\label{psis}
\end{equation}
which we denote as $\psi$'s and label by greek letters $\alpha$,
$\beta$, and so on.

The aim of the present study is to establish a proper general
formalism and to determine the parameters of the vector resonances
$\psi$ enumerated in (\ref{psis}) from a simulations fit for all
exclusive open-charm channels measured by Belle --- see
(\ref{channels}). The cornerstone of the approach used is unitarity
preserved at every stage of the data analysis. This allows one to
arrive at a selfconsistent description of the entire bulk of the data
for the measured open-charm channels, to extract parameters of the
vector $\psi$ states in the least model-dependent way, and, in
general, to establish a reliable framework for studies of various
resonances simultaneously measured in several channels.

As the contributions from the charmed-strange final states or the
three-body $D^*\bar{D}\pi$ final states are small and the only
available data from Belle have poor accuracy for these modes, these
final states are not included in the present analysis. In fact, these
minor channels can also be added into the overall fit, however this
would result in a significant increase of the number of free
parameters which will be only very weakly constrained given a very low
quality of the data in these additional channels. It is important to
notice however that, while neglecting these minor channels one
violates unitarity thus biasing the result, this bias is under control
due to the explicit unitarity of the approach used in this work.

\section{Coupled-channel model}

The traditional way to analyse the experimental data is to use
individual Breit-Wigner distributions for each peak combined with a
suitable background. It has to be noticed however that such an
approach can provide only very limited information on the states under
study. Indeed, on one hand, by analysing each reaction channel
individually, one does not exploit the full information content
provided by the measurements. Besides that the Breit-Wigner parameters
are reaction-dependent and the naive algebraic sum of the Breit-Wigner
distributions violates unitary. The last but not the least problem
with the Breit-Wigner formula is that it cannot, as a matter of
principle, describe threshold phenomena which become extremely
important for the studies above the open-flavour threshold.

We start from the amplitude $A$ in the Argand units,
\begin{equation}
S=1+2iA,
\end{equation}
and use the $K$-matrix representation for it,
\begin{equation}
A=K(1-iK)^{-1},
\end{equation}
where the $K$ matrix is Hermitian that guarantees that the amplitude
is unitary,
\begin{equation}
AA^{\dagger}=\frac{1}{2i}(A-A^{\dagger}).
\end{equation}

We assume the $K$ matrix to take the form
\begin{equation}
K_{ij}=\sum_{\alpha}G_{i\alpha}(s)\frac{1}{M_{\alpha}^2-s}G_{j\alpha}(s),
\end{equation}
where $i$ and $j$ run over hadronic channels and $\alpha$ labels the
bare $\bar{c}c$ states with the masses $M_{\alpha}$.  The form factors
$G_{i\alpha}(s)$ are defined as
\begin{equation}
G_{i\alpha}^2(s)=g_{i\alpha}^2\frac{k_i^{2l_i+1}}{\sqrt{s}}\theta(s-s_{i}),
\end{equation}
where $\theta(x)$ is the Heaviside step function, $l_i$ is the angular
momentum in the $i$-th hadronic channel and $s_{i}=(M_{1i}+M_{2i})^2$
is $i$-th threshold.

Then the amplitude $A$ reads
\begin{equation}
A_{ij}=\sum_{\alpha\beta}G_{i\alpha}(s)P_{\alpha\beta}(s)G_{j\beta}(s),
\label{amplitude}
\end{equation}
with
\begin{equation}
(P^{-1}(s))_{\alpha\beta}=(M_{\alpha}^2-s)\delta_{\alpha\beta}-i\sum_{m}G_{m\alpha}G_{m\beta}.
\label{Pm1}
\end{equation}

The couplings $g_{i\alpha}$ are defined as follows. The width $\Gamma_{i\alpha}$ for the vector $\psi_\alpha$ decaying into the $i$-th open-charm
channel $[D^{(*)}\bar{D}^{(*)}]_i$ is given by
\begin{equation}
\Gamma_{i\alpha}\equiv\Gamma(\psi_\alpha\to
[D^{(*)}\bar{D}^{(*)}]_i)=\frac{g_{i\alpha}^2}{M_\alpha^2}[p_i(M_\alpha)]^{2l_i+1},
\label{partialwidths}
\end{equation}
where
$p_i(M_\alpha)=\lambda^{1/2}(M_\alpha^2,m^2_{D_i^{(*)}},m^2_{\bar{D}_i^{(*)}})/(2M_\alpha)$
is the centre-of-mass momentum and $l_i$ is the angular momentum in
the final state. In particular, $l_i=1$ for $i=D^+D^-$,
$D^0\bar{D}^0$, $D^+D^{*-}$, $[D^{*+}D^{*-}]_{S=0}^{P}$, and
$[D^{*+}D^{*-}]_{S=2}^P$, then $l_i=2$ for the $D_2^+D^-$ final state,
and, finally, $l_i=3$ in the $[D^{*+}D^{*-}]_{S=2}^F$ channel.

In the Vector Dominance Model (VDM), the amplitude of the annihilation
process $\psi_\alpha\to\gamma^*\to e^+ e^-$ is
\begin{equation}
{\cal M}(\psi_\alpha\to e^+e^-)=\frac{g_{e\alpha}}{M_\alpha^2}\,(\bar{u}\gamma_\mu v)\epsilon^\mu,
\end{equation}
where $u$ and $v$ are the wave functions of the electron and positron, respectively, $\epsilon_\mu$ is the polarisation vector of the
$\psi_\alpha$ meson, $M_\alpha$ is its mass, $e$ is the electron
charge (we work in the Heaviside units, $\alpha=e^2/(4\pi)$), and $g_{e\alpha}$ is the $\psi$-$\gamma$ coupling
constant. Then the corresponding $\psi_\alpha$'s electronic width reads
\begin{equation}
\Gamma_{e\alpha}\equiv\Gamma(\psi_\alpha\to e^+e^-)=\frac{\alpha g_{e\alpha}^2}{3M_\alpha^3}.
\label{dielectron}
\end{equation}

It is straightforward now to proceed to the total cross section for
the annihilation process $e^+ e^- \to [D^{(*)}\bar{D}^{(*)}]_i$,
\begin{equation}
\sigma_i(s)=\frac{4\pi\alpha}{s^{5/2}}\left[p_i(s)\right]^{2l_i+1}\Bigl|\sum_{\alpha,\beta}g_{e\alpha}P_{\alpha\beta}
(s)g_ { i\beta }\Bigr|^2,
\label{cross}
\end{equation}
where the inverse propagators matrix $P^{-1}$ is defined in
(\ref{Pm1}).

One can see therefore that the set of the 16 open-charm channels (\ref{channels}) can be described with 40 parameters,
\begin{equation}
\{M_\alpha,~\Gamma_{e\alpha},~g_{i\alpha}\},\quad\alpha=\overline{1,5},\quad i=\overline{1,16},
\end{equation}
where, for convenience, we use the electronic widths
$\Gamma_{e\alpha}$ as the free parameters instead of the couplings
$g_{e\alpha}$---see their interrelation in (\ref{dielectron}).  As was explained above, due to the isospin symmetry,
model parameters for the isospin-conjugated channels  coincide with each other (except for the meson masses).

In the fitting procedure we encounter a
serious problem which reveals itself in the presence the
$D^*\bar{D}^*$ channel. Indeed, the transition amplitude acquires 3
independent contributions coming from the 3 different patterns for the
$D^*$'s helicities.  As was explained above, 2 of them correspond to
the $P$ wave with the total spin equal to 0 or 2 and the third one
corresponds to the $F$ wave with $S=2$. These 3 contributions are
treated as 3 independent channels while the available experimental
data provides only the sum of them all. Thus the fit is expected to
have a poor capability to decompose the exclusive $D^*\bar{D}^*$ cross
section into the above 3 components.

\begin{figure*}[t]
\centerline{\epsfig{file=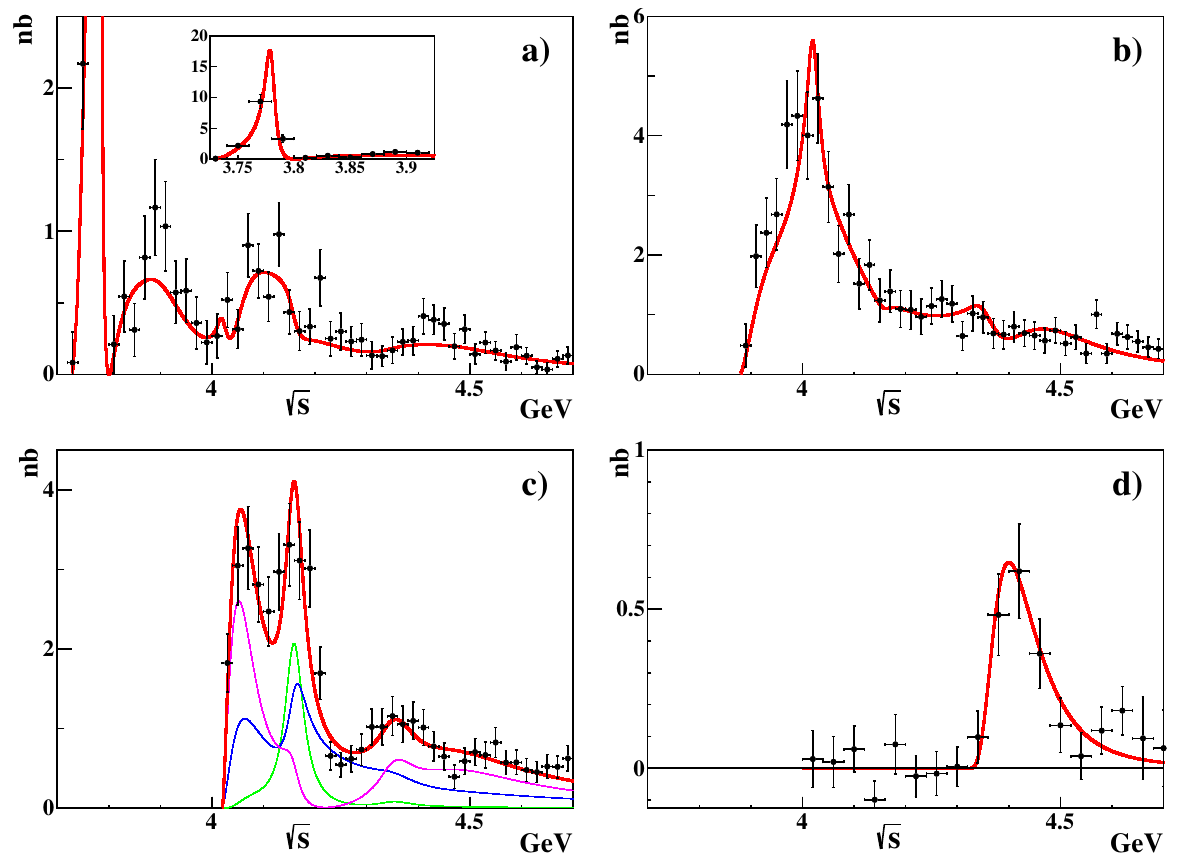,width=0.95\textwidth}}
\caption{{\bf Figure~\thefigure.}~Exclusive cross sections for the
$e^+e^-\to D\bar{D}$ $([e^+e^-\to D^+D^-]+[e^+e^-\to D^0\bar{D}^0])$ (plot (a)),
$e^+e^-\to D^+D^{*-}$ (plot (b)), $e^+e^-\to D^{*+}D^{*-}$ (plot (c)), and
$e^+e^-\to D\bar{D}\pi$ $([e^+e^-\to D^0D^-\pi^+]+[e^+e^-\to \bar{D}^0D^+\pi^-])$ (plot (d)).
In all plots, the points with the error bars represent the Belle data~\cite{durh} and the red curves show the fit
results. In plot (c), the blue, magenta and green thin curves represent the $P$-wave with $S=0$, the $P$-wave with $S=2$,
and the $F$-wave with $S=2$ contributions in $D^{*+}D^{*-}$ final state, respectively.}
\label{fig:DcDc}
\end{figure*}

To diminish the influence of this problem we employ some heavy-quark spin symmetry
(HQSS) constraints for the couplings. Namely, we use HQSS-governed $P$-wave spin-recoupling coefficients
for both $S$- and $D$-wave vectors,
\begin{eqnarray}
&&|{}^3S_1\rangle=
-\frac{1}{2\sqrt{3}}|D\bar{D}\rangle+\frac{1}{\sqrt{3}}|D\bar{D}^*\rangle_-\nonumber\\[-2mm]
\label{recouplS}\\[-2mm]
&&\hspace*{0.16\textwidth}-\frac{1}{6}|D^*\bar{D}^*
\rangle_{P0}+\frac{\sqrt{5}}{3}|D^*\bar{D}^*\rangle_{P2},\nonumber\\
&&|{}^3D_1\rangle=
\frac{\sqrt{5}}{2\sqrt{3}}|D\bar{D}\rangle+\frac{\sqrt{5}}{2\sqrt{3}}|D\bar{D}^*\rangle_-+
\frac{\sqrt{5}}{6}|D^*\bar{D}^*\rangle_{P0}\nonumber\\[-2mm]
\label{recouplD}\\[-2mm]
&&\hspace*{0.3\textwidth}-\frac{1}{6}|D^*\bar{D}^*\rangle_{P2},\nonumber
\end{eqnarray}
where $|D\bar{D}^*\rangle_-$ stands for the $C$-odd combination, and the subscripts $P0$ and $P2$ label the two different components of the
$D^*\bar{D}^*$ $P$-wave function --- see (\ref{channels}).

We assume in what follows $\psi(2S)\equiv\psi_1$,
$\psi(4040)\equiv\psi_3$, and $\psi(4415)\equiv\psi_5$ to be predominantly
$^3S_1$ states and $\psi(3770)\equiv\psi_2$ and
$\psi(4160)\equiv\psi_4$ to be predominantly  $^3D_1$ states, and relate the $P$-wave couplings in the $D^*\bar{D}^*$ channels,
\begin{eqnarray}
g_{[D^*\bar{D}^*]_{P2},\alpha}&=&-\sqrt{20}\,g_{[D^*\bar{D}^*]_{P0},\alpha},\quad \alpha=1,3,5,\nonumber\\[-2mm]
\label{relations4g}\\[-2mm]
g_{[D^*\bar{D}^*]_{P0},\alpha}&=&-\sqrt{5}\,g_{[D^*\bar{D}^*]_{P2},\alpha},\quad \alpha=2,4,\nonumber
\end{eqnarray}
so that the number of the remaining free parameters of the model is  reduced to 35.

It has to be noticed that the description of the same data sets in
terms of the naive sums of 5 Breit-Wigners in each channel would
require 15 parameters per channel (5 masses, 5 widths, 4 relative
phases, and the overall norm), that is 75 parameters in
total. Therefore the unitary approach used in this paper plus
symmetry-driven constraints reduce the number of the free parameters
by more than a factor of 2.

For the masses of the $D$ mesons involved we use the standard PDG
values \cite{Agashe:2014kda},
\begin{eqnarray}
&m_{D^0}=1864.83~\mbox{MeV},\quad m_{D^\pm}=1869.5~\mbox{MeV},&\nonumber\\
&m_{D^{*0}}=2006.85~\mbox{MeV},\quad m_{D^{*\pm}}=2010.26~\mbox{MeV},&\label{masses}\\
&m_{D_2^\pm}=2465.4~\mbox{MeV}.&\nonumber
\end{eqnarray}

\section{Fit for the experimental data}

We have checked the consistency of the model with the experimental
data and performed a simultaneous fit for the open-charm exclusive
cross sections in the interval corresponding to the known $\psi$
states. The fit minimises the function $\chi^2_{\rm exp}$ defined as a
sum of  $(\sigma_{\rm exp}-FF)^2/\delta_{\rm exp}^2$ over all Belle
experimental points. Here $FF$ is the Fitting Function for the given channel --- see  
(\ref{cross}) --- while $\sigma_{\rm exp}$ and $\delta_{\rm exp}$ are the
experimental value and the corresponding error, respectively. The
statistical and systematic experimental errors of the Belle data are
summed in squared.

The influence of the
under-threshold $\psi(2S)$ on the $D\bar{D}$ line shape is known to be important.
We therefore take into account contributions of all 5 $\psi$-resonances from (\ref{psis}) and treat the coupling constants
of all 5 states to all channels as the free parameters of the fit
(with the constrains from the HQSS imposed). The coupling constants to the
neutral and the charged modes are set equal to each other due to the
isospin symmetry and the couplings of the charge conjugated states are
also considered equal to each because of the charge parity
conservation. Meanwhile, due to a different phase space, the shapes of
the cross sections are slightly different for the neutral
and charged modes around the threshold. As the model considers only
two-body final states, we assume that the $D\bar{D}\pi$ final state is
dominated by the $D\bar{D}_2$ intermediate statethat is
consistent with the Belle study of the resonance structure in the
 $D\bar{D}\pi$ channels.

The $\psi(2S)$ resonance has the mass well below the threshold for the
considered open-charm channels. We therefore have to fix its mass and
its electronic width to the PDG values \cite{Agashe:2014kda}. In
addition, we require the $\psi(2S)$ total width to coincide with the
PDG value too.  To this end we add an auxiliary channel completely
decoupled from the other $\psi$-resonances which provides the correct
$\psi(2S)$ line shape near the pole mass. The coupling constants to
the open-charm modes for the $\psi(2S)$ are unconstrained, so that the
total number of the free parameters in the fit is eventually reduced
to 33.

We constrain the fit to converge to phenomenologically adequate
values, close to the PDG ones for the masses and the electronic
widths of the $\psi$-states. Moreover, we require the total
widths of the $\psi$-resonances to be reasonably small by adding an
extra term to the $\chi^2$,
\begin{eqnarray}
&\displaystyle\chi^2_{\rm tot}=\chi^2_{\rm 
exp}+\sum_{\alpha=1}^5\left\{\vphantom{+\left(\frac{\sum_{i=1}^{16}\Gamma_{i\alpha}}{200~\mbox{MeV}}\right)^2}
\left(\frac{M_\alpha-M^{\rm PDG}_\alpha}{50~\mbox{MeV}}\right)^2
\right.\hspace*{15mm}&\nonumber \\[-2mm]
\\[-2mm]
&\displaystyle\hspace*{15mm}
+\left(\frac{\Gamma_{e\alpha}-\Gamma^{\rm PDG}_{e\alpha}}{0.5~\mbox{MeV}}\right)^2
+\left.\left(\frac{ \sum_{i=1}^{16} \Gamma_{i\alpha}}{200~\mbox{MeV}}\right)^2
\right\},& \nonumber
\end{eqnarray}
where the indices $i$ and $\alpha$ run over the open-charm channels
and the $\psi$-resonances, as given in (\ref{channels}) and
(\ref{psis}), respectively. The formulae for the widths are given in (\ref{partialwidths}) and (\ref{dielectron}). 
This modification prevents the fit from blowing up the resonances and finding unphysical minima.

Finally, we fit 191 experimental points in all 5 open-charm channels
with the fitting function which contains 33 free parameters. It turns
out that $\chi^2_{\rm tot}$ possesses multiple local minima separated
by barriers in the parameter space. As a result, the local minima are
not continuously connected, so that the fit cannot automatically
proceed from one domain, with a bad value of $\chi^2_{\rm tot}$, to
another domain, with a better $\chi^2_{\rm tot}$.  We therefore choose
the following strategy to search for the global minimum.   We
randomly generate $10^4$ seeds for the coupling constants and
perform automatic fits. The initial values for the masses and the
electronic widths of the $\psi$-resonances are set to their
respective PDG values and then they are released as free
parameters. Then the fit with the best $\chi^2_{\rm exp}$ is
selected.

The best fit found has $\chi^2_{\rm exp}=158$, that is it is almost
perfect, given that the number of the fitted experimental points is
191 and the number of the free parameters is 33. The line shapes which
correspond to this best fit are plotted in
Figs.~\ref{fig:DcDc} and the set of parameters is listed
in Table~\ref{tab:params}. For convenience, we also quote the partial
widths of the $\psi$'s in all studied channels --- see
(\ref{partialwidths}).

A couple of concluding remarks on the fitting procedure are in order
here. As it was mentioned above, the $\psi(2S)$ tail is an important
contribution which cannot be ignored. Meanwhile, we have checked that
the shape of the $\psi(2S)$ amplitude only affects the parameters of
the $\psi(2S)$ coupling constants while the overall quality of the fit
does not change.

The second comment is that the quality of the present data does not
allow one to check how well the HQSS constraints are fulfilled in the
system under study --- see \cite{Hanhart:2015cua,Guo:2016bjq} where such
an analysis was perform for the near-threshold states $Z_b(10610)$ and
$Z_b(10650)$ in the spectrum of bottomonium. We therefore use
constraints (\ref{relations4g}) only to reduce the number of the
parameters of the fit. Meanwhile, such a check should become possible for the future, more accurate, data.

\begin{table*}[t]
\caption{{\bf Table~\thetable.}~Parameters of the $\psi$-resonances extracted from the fit for the Belle data. Parameters marked with the asterisk
are artificially fixed to the PDG values.}\label{tab:params}
\begin{center}
\begin{tabular}{|c|c|c|c|c|c|}
\hline
&$\psi_1$&$\psi_2$&$\psi_3$&$\psi_4$&$\psi_5$\\
\hline
PDG name&$\psi(2S)$&$\psi(3770)$&$\psi(4040)$&$\psi(4160)$&$\psi(4415)$\\
\hline
$M$, MeV&~~~$3686^*$~~~&$3782\pm1$&$4115\pm14 $&$4170\pm7$&$4515\pm18$\\
\hline
\multicolumn{6}{|c|}{$\vphantom{\Bigl[}$ Coupling constants $g_{i\alpha}$ ($\alpha=1\ldots 5$, $i=D\bar{D}$, $D\bar{D}^*$, {\em etc} --- see 
(\ref{channels}))}\\
\hline
$D\bar{D}$  &$3.0\pm0.3$ &$-1.8\pm0.3$ &$-0.1\pm0.1$ &$0.3\pm0.1$ &$-0.1\pm0.1$\\
\hline
$D\bar{D}^*$  &$-4.7\pm0.5$ &$-3.1\pm0.3$ &$2.4\pm0.2$ &$-0.0\pm0.7$ &$-0.7\pm0.2$\\
\hline
$[D^*\bar{D}^*]^P_{S=0}$ &$4.8\pm0.5$ &$6.9\pm0.9$ &$-0.1\pm0.2$ &$0.6\pm0.5$ &$-0.3\pm0.1$\\
\hline
$[D^*\bar{D}^*]^P_{S=2}$ &$-21.7\pm-2.3$ &$-3.1\pm-0.4$ &$0.5\pm0.9$ &$-0.3\pm-0.2$ &$1.5\pm-0.3$\\
\hline
$[D^*\bar{D}^*]^F_{S=2}$, MeV$^{-2}$  &$62.2\pm15.1$ &$-1.6\pm5.4$ &$-1.0\pm2.8$ &$8.0\pm1.4$ &$0.2\pm0.6$\\
\hline
$D_2\bar{D}$, MeV$^{-1}$  &$-8.2\pm29.3$ &$25.2\pm7.7$ &$-23.5\pm3.3$ &$-1.0\pm7.4$ &$-1.5\pm1.4$\\
\hline
\multicolumn{6}{|c|}{$\vphantom{\Bigl[}$ Partial decay widths $\Gamma_{i\alpha}$, MeV}\\
\hline
$e^+e^-$ &$2.354^*$&$0.2\pm0.0$&$1.6\pm0.3$&$0.7\pm0.4$&$1.4\pm0.3$\\
\hline
$D^+D^-$& -&$5.6\pm1.7$&$0.4\pm0.8$&$4.3\pm2.6$&$0.5\pm1.0$\\
\hline
$D^0\bar{D}^0$& -&$7.5\pm2.2$&$0.4\pm0.8$&$4.5\pm2.7$&$0.5\pm1.0$\\
\hline
$D^+D^{*-}$& - & - &$110.7\pm23.5$&$0.0\pm0.5$&$32.8\pm17.4$\\
\hline
$[D^*\bar{D}^*]^P_{S=0}$& - & - &$0.1\pm0.2$&$3.6\pm6.5$&$5.9\pm2.6$\\
\hline
$[D^*\bar{D}^*]^P_{S=2}$& - & - &$1.2\pm6.8$&$0.7\pm0.3$&$118.0\pm729.4$\\
\hline
$[D^*\bar{D}^*]^F_{S=2}$& - & - &$0.2\pm1.0$&$58.6\pm22.9$&$2.3\pm14.2$\\
\hline
$D_2^+D^-$& - & - & - & - &$11.7\pm21.1$\\
\hline
\end{tabular}
\end{center}
\end{table*}

\section{Summary}

In this paper we used a coupled-channel approach to perform a
simultaneous fit for the data in the major open-charm channels
measured by Belle in a wide energy range $\sqrt{s}=3.7\div
4.7$~GeV. The main advantage of the approach used here as compared to
previous works is that unitarity of the amplitude is under control at
every stage of the data analysis. In particular, unitarity is
preserved up to the minor contributions of the neglected strange-charm
and many-body channels (we have checked that, if the model is extended
to include the $D_s^+D_s^-$ channel, then the overall description of
the data improves only marginally). This allows one to link tightly
the parameters in different channels and, as a result, to considerably
reduce the number of free parameters of the fit. The main conclusion
of the paper is that the suggested method is indeed able to explain
all data sets simultaneously and provides a very good overall
description of the line shapes. The presence of multiple local minima
of the $\chi^2$ should be attributed to a relatively low quality of
the present data, that makes a straightforward interpretation of the
parameters of the $\psi$-resonances extracted from the fit
questionable. Meanwhile, the situation may improve considerably when the next update for the data in the open-charm channels
appears.
\medskip

This work is supported by the Russian Science Foundation (Grant
No. 15-12-30014).


\begin{thebibliography}{99}

\bibitem{MarkI:cs_4415} J. Siegrist, G. S. Abrams, A. M. Boyarski {\it et al.} (MARK-I Collaboration), Phys.\ Rev.\ Lett. {\bf 36}, 700 (1976).

\bibitem{MarkI:cs_3770} P. A. Rapidis, B. Gobbi, D. L{\"u}ke {\it et al.} (MARK-I Collaboration), Phys.\ Rev.\ Lett. {\bf 39}, 526 (1977),
Erratum-ibid. {\bf 39}, 974 (1977).

\bibitem{DASP:cs} R. Brandelik, W. Braunschweig, H.-U. Martyn {\it et al.} (DASP Collaboration), Phys.\ Lett.\ B {\bf 76}, 361 (1978).

\bibitem{DELCO:cs_3770} W. Bacino, A. Baumgarten, L. Birkwood {\it et al.} (DELCO Collaboration), Phys.\ Rev.\ Lett. {\bf 40}, 671 (1978).

\bibitem{MarkII:cs_3770} R. H. Schindler, J. L. Siegrist, M. S. Alam {\it et al.} (MARK-II Collaboration), Phys.\ Rev.\ D {\bf 21}, 2716 (1980).

\bibitem{seth:fit} K.~K.~Seth, Phys.\ Rev.\ D {\bf 72}, 017501 (2005).

\bibitem{cb:cs} A. Osterheld, R. Hofstadter, R. Horisberger {\it et al.}  (Crystal Ball Collaboration), SLAC-PUB-4160, (1986).

\bibitem{bes:cs} J. Z. Bai , Y. Ban, J. G. Bian {\it et al.} (BES Collaboration),
Phys.\ Rev.\ Lett. {\bf 88}, 101802 (2002).

\bibitem{bes:fit} M. Ablikim, J.Z. Bai, Y. Ban {\it et al.} (BES Collaboration),
Phys.\ Lett.\ B {\bf 660}, 315 (2008).

\bibitem{babar:y4260} B. Aubert, R. Barate, D. Boutigny {\it et al.} (BaBar Collaboration),
Phys.\ Rev.\ Lett. {\bf 95}, 142001 (2005).

\bibitem{babar:y4260_08} B. Aubert, M. Bona, Y. Karyotakis {\it et al.} (BaBar Collaboration),
arXiv:0808.1543, (2008).

\bibitem{babar:y4360} B. Aubert, R. Barate, M. Bona, {\it et al.} (BaBar Collaboration),
Phys.\ Rev.\ Lett. {\bf 98}, 212001 (2007).

\bibitem{belle:y4260:abe} K. Abe, K. Abe, I. Adachi {\it et al.} (Belle Collaboration),
hep-ex/0612006.

\bibitem{belle:y4260:yuan} C. Z. Yuan, C. P. Shen, P. Wang {\it et al.} (Belle Collaboration),
Phys.\ Rev.\ Lett. {\bf 99}, 182004 (2007).

\bibitem{belle:y4360} X. L. Wang , C. Z. Yuan, C. P. Shen {\it et al.} (Belle Collaboration),
Phys.\ Rev.\ Lett. {\bf 99}, 142002 (2007).

\bibitem{Wang:2014hta} X. L. Wang, C. Z. Yuan, C. P. Shen {\it et al.} (Belle Collaboration),
Phys.\ Rev.\ D {\bf 91}, 112007 (2015).

\bibitem{Liu:2013dau} Z. Q. Liu, C. P. Shen, C. Z. Yuan {\it et al.} (Belle Collaboration),
Phys.\ Rev.\ Lett.\ {\bf 110}, 252002 (2013).

\bibitem{belle:lala} G. Pakhlova, I. Adachi, H. Aihara {\it et al.} (Belle Collaboration),
Phys.\ Rev.\ Lett. {\bf 101}, 172001 (2008).

\bibitem{cleo:eichten} E.~Eichten, K.~Gottfried, T.~Kinoshita,
K.~D.~Lane, and T.~M.~Yan, Phys.\ Rev.\ D {\bf 21}, 203 (1980).

\bibitem{belle:dd} G. Pakhlova, I. Adachi, H. Aihara {\it et al.} (Belle Collaboration),
Phys.\ Rev.\ D {\bf 77}, 011103 (2008).

\bibitem{belle:ddst} K. Abe, I. Adachi, H. Aihara {\it et al.} (Belle Collaboration),
Phys.\ Rev.\ Lett. {\bf 98}, 092001 (2007).

\bibitem{belle:ddpi} G. Pakhlova, I. Adachi, H. Aihara {\it et al.} (Belle Collaboration),
Phys.\ Rev.\ Lett. {\bf 100}, 062001 (2008).

\bibitem{belle:ddstpi} G. Pakhlova, I. Adachi, H. Aihara {\it et al.} (Belle Collaboration),
Phys.\ Rev.\ D {\bf 80}, 091101 (2009).

\bibitem{belle:dsds} G. Pakhlova, I. Adachi, H. Aihara {\it et al.} (Belle Collaboration),
Phys.\ Rev.\ D {\bf 83}, 011101 (2011).

\bibitem{babar:dd} B. Aubert, R. Barate, M. Bona {\it et al.} (BaBar Collaboration),
Phys.\ Rev.\ D {\bf 76}, 111105 (2007).

\bibitem{babar:ddst} B. Aubert, Y. Karyotakis, J. P. Lees {\it et al.} (BaBar Collaboration),
Phys.\ Rev.\ D {\bf 79}, 092001 (2009).

\bibitem{babar:dsds} P. A. Sanchez, J. P. Lees, V. Poireau {\it et al.} (BaBar Collaboration),
Phys. Rev. D {\bf 82}, 052004 (2010).

\bibitem{cleo:cs} D. Cronin-Hennessy , K. Y. Gao, J. Hietala {\it et al.} (CLEO Collaboration),
Phys. Rev. D {\bf 80}, 072001 (2009).

\bibitem{Agashe:2014kda} K.A. Olive, K. Agashe, C. Amsler {\it et al.} (Particle Data Group Collaboration),
Chin.\ Phys.\ C {\bf 38}, 090001 (2014).

\bibitem{Li:2009pw} H.~B.~Li, X.~S.~Qin and M.~Z.~Yang,
Phys.\ Rev.\ D {\bf 81}, 011501 (2010).

\bibitem{Cao:2014vca} X.~Cao and H.~Lenske,
arXiv:1408.5600 [nucl-th].

\bibitem{Cao:2014qna} X.~Cao and H.~Lenske,
arXiv:1410.1375 [nucl-th].

\bibitem{Limphirat:2013jga} A.~Limphirat, W.~Sreethawong, K.~Khosonthongkee, and Y.~Yan,
Phys.\ Rev.\ D {\bf 89}, 054030 (2014).

\bibitem{Achasov:2012ss} N.~N.~Achasov and G.~N.~Shestakov,
Phys.\ Rev.\ D {\bf 86}, 114013 (2012).

\bibitem{Achasov:2013lwk} N.~N.~Achasov and G.~N.~Shestakov,
Phys.\ Rev.\ D {\bf 87}, 057502 (2013).

\bibitem{Zhang:2009gy} Y.~J.~Zhang, Q.~Zhao,
Phys.\ Rev.\ D {\bf 81}, 034011 (2010).

\bibitem{Zhang:2010zv} Y.~J.~Zhang and Q.~Zhao,
Phys.\ Rev.\ D {\bf 81}, 074016 (2010).

\bibitem{Segovia:2011zza} J.~Segovia, D.~R.~Entem, and F.~Fernandez,
Phys.\ Rev.\ D {\bf 83}, 114018 (2011).

\bibitem{Segovia:2013wma} J.~Segovia, D.~R.~Entem, F.~Fernandez, and E.~Hernandez,
Int.\ J.\ Mod.\ Phys.\ E {\bf 22}, 1330026 (2013).

\bibitem{vanBeveren:2010mg} E.~van Beveren, G.~Rupp, and J.~Segovia,
Phys.\ Rev.\ Lett. {\bf 105}, 102001 (2010).

\bibitem{durh} Durham database,\\
http://durpdg.dur.ac.uk/review/rsig/BELLE.shtml

\bibitem{Hanhart:2015cua} C.~Hanhart, Yu.~S.~Kalashnikova, P.~Matuschek, R.~V.~Mizuk, A.~V.~Nefediev, and Q.~Wang,
Phys.\ Rev.\ Lett.\  {\bf 115}, 202001 (2015).

\bibitem{Guo:2016bjq} F.-K.~Guo, C.~Hanhart, Yu.~S.~Kalashnikova, P.~Matuschek, R.~V.~Mizuk, A.~V.~Nefediev, Q.~Wang, and J.-L.~Wynen,
Phys.\ Rev.\ D {\bf 93}, 074031 (2016).

\end{thebibliography}
\end{document}